# Perfect Seismic Wave Absorbers for the 8[th] Magnitude Earthquakes

Liang Sun[1, (a)], Feilong Xu[2, 3, (a)], Zhen Liang*[3], X. D. He[2, 4], Z. Yang*[1, 2]

1, Department of Physics, The Hong Kong University of Science and Technology, Clearwater Bay, Kowloon, Hong Kong, The People's Republic of China

2, Xichuang Advanced Materials Research Institute, 6[th] Floor, Phase II, Entrepreneurship Center for Overseas Graduates, 29 South Gaoxin Circle, Nanshan District, Shen Zhen, Guangdong, The People's Republic of China

3, Harbin Institute of Technology (Shenzhen), School of materials science and engineering, Shenzhen, Guangdong, The People's Republic of China

4, Center for Composite Materials and Structures, School of Astronautics, Harbin Institute of Technology, Harbin 150080, The People's Republic of China

a) Made equal contributions. Names are listed alphabetically.

* Corresponding authors

## Abstract

Most modern buildings and infrastructures are not designed to resist 8[th] magnitude earthquakes, for which studies on phononic crystals and local resonators reported so far are not particularly suited. We report a type of all-metallic decorated membrane resonators as deep subwavelength vibration dampers, which are modeled as meta-soil, that can block and even totally absorb underground seismic waves up to 8[th] magnitude earthquakes. Transmission attenuation exceeding 20 dB and absorption up to 98 % are numerically demonstrated for 1 Hz Rayleigh waves by a dozen rows of vertical underground wells filled with the meta-soil. A scaling law in the same form as the mass density law for airborne acoustic waves has been analytically derived and numerically verified.

## Introduction

Earthquakes are one of the most devastating natural disasters to human lives and infrastructures. Numerous researches and developments have been carried out in the past few decades to mitigate the damage to buildings and other infrastructures when seismic waves reach and shake their foundations, while little work had been done, until in recent years, to attenuate the seismic waves before reaching the foundations [1]. Modern buildings are designed to resist earthquakes up to 7[th] magnitude [1]. Therefore, the real need to protect modern cities against earthquake rests upon means that target at the 8[th] magnitude earthquakes. The vibration spectra of most earthquakes centered at certain frequencies, which becomes lower for higher magnitude earthquakes [1]. The very low central frequency (~ 1 Hz) of the 8[th] magnitude earthquakes makes the mitigation extremely difficult. For epic centers some distance away, the most destructive waves are the Rayleigh waves, while the bulk pressure and shear waves are significantly attenuated by geometric effect [1]. Looking no further than the damages caused by the two earthquakes in 2016, one must admit that there is still a long way to go in human struggle against earthquakes, which desperately calls for breakthrough technologies.

Phononic crystals [2 − 5] and local resonator metamaterials [6 − 17] offer a viable alternative to the early attempts of using simple structures such as ditches and so forth to block the underground seismic waves [18]. Periodically positioned hollow cylinders, which serve as individual scattering centers for Brag reflections, were dug into the ground to form a phononic crystal with the band gap frequency around 50 Hz [3, 4]. More recently, phononic crystals formed by square steel piles filled with concrete embedded periodically in soil were theoretically investigated. These investigations, while providing a proof of concept for the viability of using phononic crystals to block the underground seismic waves, also reveal two major drawbacks of the approach when applied to the 8[th] magnitude earthquakes. The first is that the period of the crystals must be comparable to the wavelength of the seismic waves, which is of the order of 100 meters even in soft soil for the 8[th] magnitude earthquakes [1]. The second is that the wave attenuation by phononic crystals depends very much on the perfection of the crystals. A large and highly uniform ground must be available for a crystal array to be implanted. Any irregularities and uneven distribution of soil, rock, etc., could significantly weaken the performance. Likewise, change of moisture in the soil by rain and drought could significantly change the wave speed and the gap frequency of the phononic crystals. Therefore, while it may be viable to use phononic crystals for the lower magnitude earthquakes or other vibration sources at much higher frequency, they are not the best candidate, to say the least, for the 8[th] magnitude earthquakes.

The local resonators acoustic metamaterials [7 − 17], which are essentially made of an oscillator of certain mass connected by an elastic spring to an anchor [6], offer an alternative means. By its very nature, each individual local resonator in a resonator array essentially functions independently of the others nearby, and their collective effect is due more to the cumulative contributions of individual ones than to the careful arrangement of their positions in order to facilitate Brag reflections. Since the interference among the waves scattered by the resonators no longer plays an essential role, the geometric size of the resonator arrays for earthquake protection could be drastically reduced. The resonant frequencies of the local resonators are the essential element in determining the frequencies of the seismic waves to be attenuated. A chain of local resonators, each consists of a sphere rolling on a cycloidal surface about 10 cm in size, were shown theoretically to filter the shear waves in solids [7]. Although the resonant frequency of 1.62 Hz was close to the central frequency of the 8[th] magnitude earthquakes, the ground displacement of the order of 10 cm in such earthquakes could easily 'spill over' the rolling spheres out of the cycloidal surface or hitting the walls, rendering the local resonators dysfunctional. Arrays of local resonators similar in design to the coated spheres [6] and essentially function like a mass-on-spring oscillator system, such as cylindrical tubes containing a resonator suspended by soft bearings [8], iron spheres connected to concrete blocks via iron or rubber ligaments [9], cross-shaped steel cavities or cylinders either being hollow or coated with rubber about 10 meters in size [10], and cylindrical mass suspended by elastomeric springs within a concrete cage [11], were shown theoretically to have good, and in some cases excellent attenuation of seismic waves. A scaled experimental investigation [11] found qualitative agreement between the theoretical predictions and the experimental outcomes. To be noted in all these theoretical investigations was the exclusion of dissipation in the elastic bearing materials, which essentially made these resonators nearly dissipation-free, or having very large $Q$-factor. The direct consequence of such exclusion is that a small disturbance to the anchor at the resonant frequency would be amplified several hundred times in the displacement of the oscillator. As the anchor of each local resonator is firmly attached to the ground, a 10 cm ground displacement from the 8[th] magnitude earthquakes would lead to over 10 meters of displacement of the oscillator in a local resonator, which none of the designs reported so far could have accommodated.

Hinted by geophysical experiments, it was discovered that local resonators on the soil surface, such as trees, could attenuate Rayleigh waves [12]. Further theoretical studies on more systematic designs of pillar-like single resonant frequency [13] and multiple resonant frequencies [14] structures indicate that they could effectively convert the surface Rayleigh waves into bulk-like shear waves and scatter them towards the underground abyss. Similar effects are found with thick columns of steel planted on the soil surface [15, 16]. A set of columns of different lengths arranged in a wedge could serve as a broadband wave converter that turns Rayleigh waves into bulk-like waves [17]. Similar effect could also be realized by a large slope [19], but the reduction in vibration displacement is very limited.

So far, all the investigations based on local resonators are qualitative and for proof of concept only, because no realistic dissipation was considered in the investigations. To move forward, it is time to carry out rigorous quantitative studies using realistic materials parameters. Most modern buildings are designed to resist up to the 7[th] magnitude earthquakes [1], so the real need for modern cities are barriers that can reduce the seismic waves from an 8[th] magnitude earthquake to 7[th] magnitude or below. Accordingly, we propose four major criteria for such purpose. (1) The attenuation frequency band should be around 1 Hz, and approximately spread from 0.5 to 2 Hz; (2) Sufficient space must be made available for the oscillators in the local resonator devices to follow the ground displacement in soil up to 10 cm; (3) The devices must maintain intimate contact with the surrounding ground during the entire earthquake which usually lasts for about 100 seconds; (4) Collectively the devices could form a barrier with at least 20 dB in transmission attenuation so that the buildings behind the barrier could survive an 8[th] magnitude earthquake.

We now examine the works reported so far [2 − 5, 7 − 17] according to these criteria. While the first criterion could be met by simple upscaling of the dimensions and the oscillator mass of the local resonators to reach the resonant frequency of 1 Hz, the second criterion has been ignored in all the previous works [7 − 17]. As dissipation was absent, the Q-factors of these resonators would be very large (at least ~100). As the ground displacement of 10 cm would be magnified by more than 100 times for the oscillator in a local resonator at resonance, none of the buried structures studied by far [7 − 12] could accommodate the maximum displacement of the oscillator which could be tens of meters. The pillar-like resonators planted on soil surface do not have movement limitation above ground [13 − 17]. However, the pillar resonators for 1 Hz must be several meters in lateral dimension, over 10 meters high and weigh many tons. The massive pillars could easily be detached from the soil base during the first few moments of ground motion and fall down, causing the destruction of the barriers before the main shock comes. As the structural elements of the seismic barriers in phononic crystals [2 − 5] and most local resonators [8 − 11, 13 − 17] are in the form of massive and rigid solid blocks, they could easily be detached to the surround soil after the first few moments of ground motion. Therefore, a practical barrier must maintain its intimate contact with the soil throughout an earthquake.

Another shortcoming in ignoring the dissipation in the local resonators is the inevitable over estimations of their wave attenuation power. As will be shown below, a row of local resonators, with nearly unlimited Q-factors, could attenuate waves with incredible power, up to 150 dB as in Ref. 8. Therefore, to meet criterion-4, one must use realistic dissipation values in the theoretical modeling to determine what is really needed to reduce the seismic waves at 1 Hz by 20 dB.

In this paper, we present a class of compact and light weight local resonators and the means of deploying these resonator devices to block and even totally absorb the seismic waves of the 8[th] magnitude earthquakes, while satisfying all four criteria stated above. In addition, we report a scaling law that connects the attenuation frequency with the oscillator mass of the local resonator, which is in the same form as the acoustic mass density law [20].

**Underground dampers**

The tuned mass dampers (TMD's), or the dynamic vibration absorbers, were invented nearly 100 year ago to damp vibration at specific frequencies [21, 22]. A single-degree-of-freedom generic version is shown in the left half of Fig. 1(a). As far as the anchor of the damper is concerned, the dynamic effect of the rest of the damper can fully be described by the effective dynamic mass attached to the anchor given by

$$\tilde{m} \equiv m_0 f(\omega) = m_0 \frac{(\omega_0^2 + i\omega\omega_0 / Q)}{\omega_0^2 - \omega^2 + i\omega\omega_0 / Q} \qquad (1),$$

where $\omega_0$ is the resonant frequency and $Q$ the quality factor. The derivation of Eq. (1) is given in the Supplementary Information (SI). When the damper is attached to a primary structure via the anchor with mass $M_A$, the primary structure would simply 'see' a dynamic mass $\tilde{m} + M_A$ if the physical size of the damper can be ignored, which is valid for deep subwavelength devices. Maximum damping effect to the vibration of the primary structure is reached in the frequency range near the resonant frequency $\omega_0$ [21, 22].

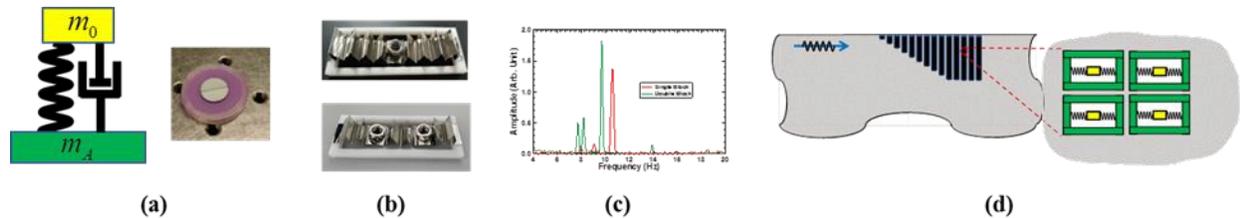

Figure 1 (a) A generic tuned mass damper (left half) and a photo of a DMR (right half); (b) Photos of a single mass block damper (upper half) and double mass block damper (lower half); (c) The corresponding free vibration spectra for the dampers in (b). (d) A schematics of wells filled with the dampers of the type shown in (b) as underground seismic wave barriers and absorbers.

It has been established recently that decorated membrane resonators (DMR's) [23] can serve as compact and light weight vibration dampers with single or multiple working frequencies [24 – 26]. A typical one is shown in the right half of Fig. 1(a), which consists of a rubber membrane (purple in color) mounted on a translucent plastic circular ring about 25 mm in diameter and decorated by a rigid metallic disk. The elastic membrane serves as both the very soft spring and the dissipating dashpot. Using membranes to provide the restoring forces, DMR's with resonant (working) frequency $\omega_0$ around 100 Hz while having total weight about 1.5 g have been demonstrated [25]. Furthermore, they can be effectively treated as equivalent TMD's for their damping effect on a primary structure [25].

Due to the limited life span of plastics and rubbers, all metallic DMR's are preferred if they are to be deployed underground for decades. A photo of an all-metallic DMR-like damper is shown in the upper half of Fig. 1(b). The damper is 120 mm long and its height is 10 mm, and its resonant frequency $\omega_0$ is at 10 Hz, as can be seen in the free vibration spectrum (red curve) in Fig. 1(c). In this DMR-like damper the soft rubber membrane in the original DMR [23] was replaced by folded stainless steel sheet about 0.1 mm in thickness. A M6-size nut (2 g in mass) served as the mass block oscillator of the DMR-like damper. A two-mass-block version shown in the lower half of Fig. 1(b) has several vibration modes around 10 Hz, as can be seen in the spectrum (green curve) in Fig. 1(c). The folded corners of the steel sheet are the regions of high elastic energy concentration because of the large curvature that provide energy dissipation much larger than the flat sheets. That is the reason why the $Q$-factor of these DMR-like dampers in Fig. 1(b) at 10 Hz is only 30, as compared to over 1000 if flat sheets are used instead. The details about the device and the measurements are presented in SI. Using the design parameters extracted from these experimental results, we estimated that a DMR-like damper with a 20 g single mass block and a length around 300 mm for the folded steel sheet would reach the resonant frequency of 1 Hz. The required length of the folded sheet would be 560 mm if the mass block remains at 2 g.

The compact DMR-like dampers make it possible to deploy large number of them underground to dissipate seismic wave energy. While in principle there are many ways to deploy the dampers, in this paper we only investigate the cases in which the dampers are installed in rows of vertical wells of certain length dug into homogenous soil ground, as shown schematically in Fig. 1(d). The dampers, each in a sealed rigid box, are placed in the wells at certain interval, and the remaining space of the wells is filled with the same soil as the surrounding one. As each DMR-like damper takes large amount of fine grids and storage memory to compute numerically, it becomes unbearable to numerically simulate the effects on seismic waves by millions of these devices deployed underground in a non-periodic fashion. Following our earlier works [24, 25], in the present simulations we replaced each DMR-like damper by a generic TMD with fictive spring and dashpot. They move as a point mass with the surrounding soil, like a small (deep subwavelength) piece of meta-rock with the relevant dynamic effective mass given by Eq. (1). To further reduce the amount of computation, the meta-rocks filling the vertical wells are homogenized as uniform elastic medium (meta-soil) fully compliant with the surrounding soil but with a dynamic effective mass density which is equivalent to the total dynamic mass of the DMR-like dampers per unit volume.

**Simplified meta-surface**

To understand the essence of damping the seismic waves by compact underground dampers, we first derive the analytic formula for the transmission and the reflection of Rayleigh waves by an infinitely large meta-surface which is fully compliant with the surrounding soil and with a given dynamic areal mass density.

Consider a Rayleigh wave propagating along the $X$-axis, and take $Z$ as the vertical axis and $Y$ as the lateral axis. The displacement in the $Z$-direction is $w = g(z)e^{ikx}$, where $g(z)$ is the depth profile of the Rayleigh wave amplitude [1]. On the incident side, the total displacement is $w_1 = w_I + w_R = g(z)e^{ikx} + rg(z)e^{-ikx}$. The displacement on the transmit side is $w_2 = t \cdot g(z)e^{ikx}$. The time harmonic function for each displacement field has been ignored. Here $t$ is the transmission coefficient and $r$ is the reflection coefficient. $TL \equiv -20\log(t)$ is then the transmission attenuation in dB.

The boundary conditions at the meta-surface ($x = 0$) are (i) the displacements being continuous, i. e.,

$$1 + r = t \qquad (2a),$$

and (ii) the Newton's second law for a unit area of the meta-surface

$$G \frac{\partial w_1}{\partial x}\Big|_{x=0} - G \frac{\partial w_2}{\partial x}\Big|_{x=0} = -\tilde{m}\omega^2 w_2(0) \qquad (2b),$$

where $G$ is the shear modulus, and $\tilde{m}$ is the dynamic areal mass density of the meta-surface given by Eq. (1). Solving Eqs. (2) we have

$$t = \frac{2}{2+iK} \qquad (3a),$$

$$r = \frac{iK}{2+iK} \qquad (3b),$$

where

$$K \equiv \frac{\tilde{m}\omega v}{G} \qquad (3c).$$

Here $v$ is the wave velocity. It can be readily shown that the absorption $A \equiv 1 - |t|^2 - |r|^2 = 0$ if $K$ is real, i. e., a block mass alone, like a piece of real rock, does not dissipate energy. Similar results can be obtained for the $X$-component of the Rayleigh waves when $G$ is replaced by the Young's modulus $Y$ in Eq. (3c).

It is easily seen from Eq. (1) that at the resonant frequency the dynamic mass is $\tilde{m} = iQm_0$. Without taking into account the realistic dissipation, the peak mass value of $\tilde{m}$ could be very large, leading to very low transmission even with only a small value of oscillator mass $m_0$. In other words, the wave attenuation of local resonators could be drastically over-estimated if dissipation is ignored.

The factor $\tilde{m}\omega$ in Eq. (3c) implies a scaling law for the dependence of the wave attenuation of the meta-surface on frequency and the areal mass density, as $v$ and $G$ are almost independent of the wave frequency. It indicates that lower dynamic mass is required to provide the same attenuation for the higher frequency waves. Therefore, lower magnitude earthquakes with higher central frequencies than the 8[th] magnitude ones could readily be attenuated by the dampers for the 8[th] magnitude ones if the dampers also have higher order working frequencies, which are characteristics of multiple-mass DMR-like dampers such as the ones shown in Fig. 1(c) and in Ref. 25. In addition, as the wave speed scales like $\sqrt{G}$, $K$ scales like $1/\sqrt{G}$, i. e, softer ground with slower wave speed is preferred for the buried dampers to attenuate seismic waves.

**Simulation conditions**

We now proceed to the numerical simulations. Consider an infinitely long row of equally spaced and identical cylindrical wells dug into the soil ground along the Y-axis of the coordinate system. The mass density of the soil is $\rho = 2000$ kg/m$^3$, the Young's modulus is $E = 100$ MPa, and the Poisson ratio is $\upsilon = 0.3$. Given the soil parameters, the Rayleigh wave velocity [1] is 128.6 m/s. The wall of the wells is covered by a thin metallic shell to maintain the integrity of the well while fully compliant with the ground motion. The bases of the dampers anchored on the well sidewall are assumed to be moving in complete compliance with the motion of the nearby soil in the event of ground vibration. Therefore, the dampers can be homogenized as meta-soil which is fully compliant with the motion of soil outside the wells. The dynamic effective mass density of the meta-soil is given by Eq. (1), with the quality factor being fixed at $Q = 10$ throughout the paper. Such $Q$-factor value is compatible to the DMR-like dampers we have reported above. The ground motion will be amplified 10 times in the TMD if real TMD's are deployed, i. e., the peak displacement of the oscillators would be up to 1 meter when the ground motion is 10 cm for the 8[th] magnitude earthquakes [1], which could be easily accommodated by the wells with diameters of 5 meters. In the case for the meta-soil, there is no limitation on the vibration amplitude, as the displacement of the meta-soil follows the surrounding real soil. The Rayleigh waves is numerically simulated using COMSOL Multiphysics in a rectangular segment of soil 6000 meters along the $X$-axis, 400 meters deep along the $Z$-axis, and 10 meters wide along the $Y$-axis as shown in Fig. 2(a). A line source parallel to the $Y$-axis is placed at $x = -4000$ m, as marked by the red arrow. The $Z$-axis of the coordinate system is parallel to the well axes and perpendicular to the horizontal ground surface, with $z = 0$ being the soil surface. The wave propagation is along the $X$-axis. At both ends of the segment along the $X$-axis are non-reflection boundaries. The diameter of the wells is fixed at 5 meters throughout the paper. The center-to-center separation between two adjacent wells is fixed at 10 meters throughout the paper. As the well diameter and the separation are much shorter than the seismic wavelength at 1 Hz, the interference effect among the scattered waves by the wells is ignored, and periodic boundary condition along the $Y$-axis with the period of 10 meters is imposed. The first row of wells will be at $x = 0$ as marked by the white arrow, if wells are introduced in the simulations. In the cases where there are several rows of wells, additional rows are placed behind the first row (to the right side in the figure) at 10 meter interval.

The line source generates longitudinal, shear, and Rayleigh waves. Figure 2(b) depicts the wave intensity distribution in the $X$-$Z$ plane without the wells. It can be seen that by the time the waves reach at $x = 0$ after traveling by over 3000 meters, the two types of bulk waves spread out mostly into deep underground abyss, and become negligible when compared to the Rayleigh waves, which propagate only along the soil surface. The simulation results confirm that the surface waves maintain all the characteristics of Rayleigh waves [1], such as the wave speed predicted by the materials parameters of the soil, the X-to-Z vibration amplitude ratio, and the depth profile of the amplitudes $g(z)$.

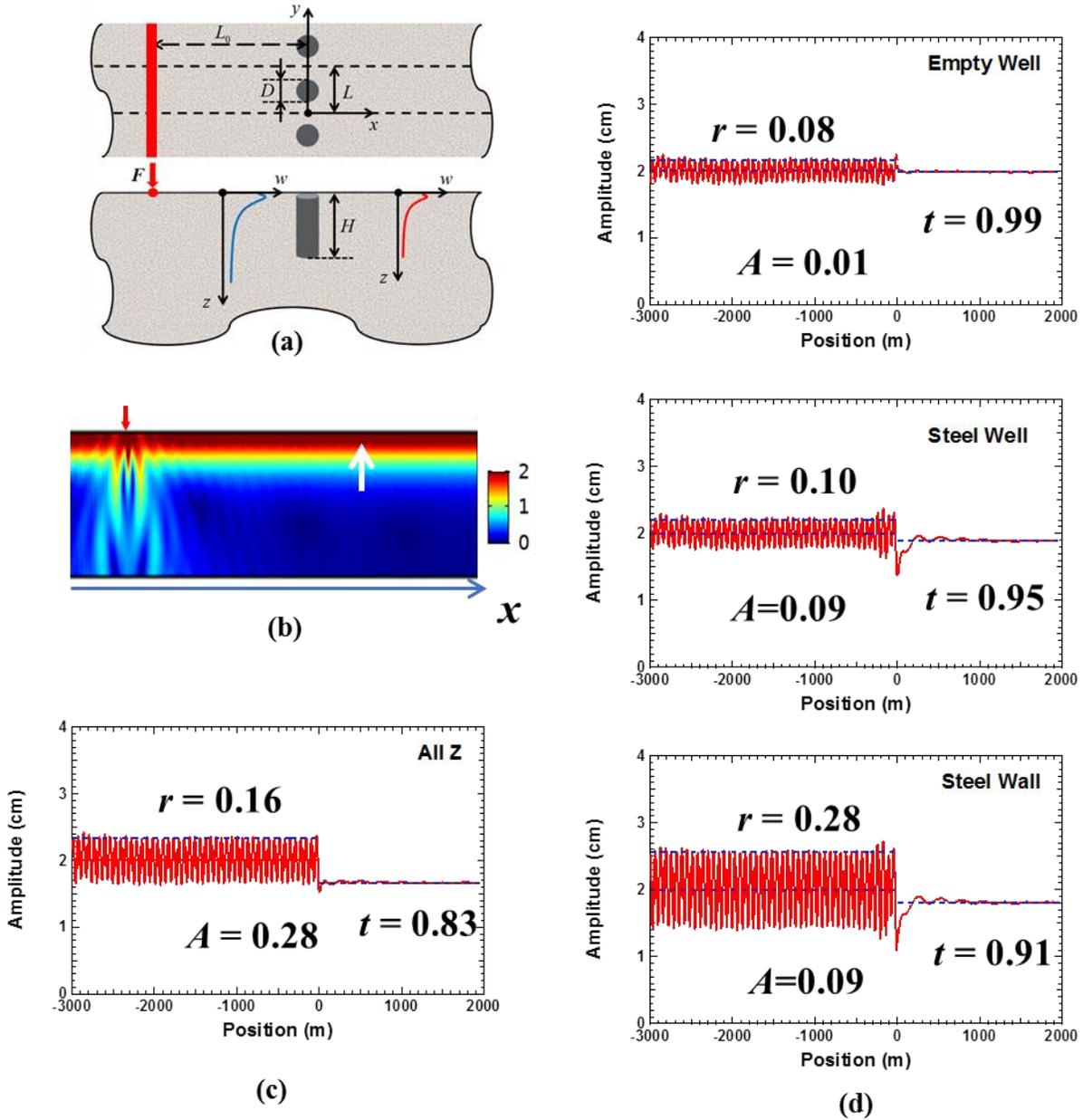

Figure 2 (a) Schematics of the soil block for the numerical simulations; (b) The color coded intensity distribution of the waves emitted by the line source marked by the red arrow. The white arrow marks the position of the first row of wells if they are present; (c) The Rayleigh wave intensity distribution at the soil surface for a row of wells filled with meta-soil for Z-direction ground vibration. (d) The Rayleigh wave intensity distribution at the soil surface for a row of empty wells (top), a row of wells filled with solid steel (middle), and a solid wall of steel (bottom).

## Case studies

### (A)    Polarization dependence

We first examine the cases with wells filled with meta-soil. In the later part of the paper we will compare the cases of meta-soil wells with the wells containing generic TMD's which provide the same average dynamic mass density, and show that the two cases are consistent with one another. The effective mass density tensor for the meta-soil then takes the form $\ddot{\rho} = \begin{pmatrix} \rho_X(\omega) & 0 \\ 0 & \rho_Z(\omega) \end{pmatrix}$, as there is no ground motion in the Y-direction for Rayleigh waves. In the mass tensor we take $\rho_Z(\omega) = \rho_0 f(\omega)$, $\rho_X(\omega) = \rho_0 = 1000$ Kg/m$^3$ unless otherwise specified, where $f(\omega)$ is given in Eq. (1) with $\omega_0 = 1$ Hz and $Q = 10$. Such value of inert mass density $\rho_0$ corresponds to the cases where about 1/8 of the space in the wells is occupied by steel oscillators. The mass density is anisotropic in that the homogenized TMD's in the wells only damp the vibration in the $Z$-direction. In the $X$-direction the meta-soil behaves like inert mass. The top graph in Fig 2(c) depicts the wave intensity distribution on the soil surface when a single row of wells with length of 100 meters are placed in front of the incoming Rayleigh waves at $\omega = 1$ Hz. The reflected waves and the transmitted waves maintain the Rayleigh wave characteristics, so unless specified otherwise, in the remaining part of the paper only the vertical motion on the soil surface is taken as the full representation of the Rayleigh wave displacement. On the incident side of the wells a standing wave pattern can be clearly seen, which is due to the reflection of the wells. Below the bottom of the wells are traces of scattered waves propagating into the underground abyss. Using the formula $\dfrac{I_{Max}}{I_{Min}} = \dfrac{1+r}{1-r}$, the reflection coefficient estimated from the standing wave pattern is 0.16, while the transmission is 0.83, and the apparent absorption is 0.28, which includes the contribution from both the dissipation by the meta-soil and the scattered waves into the underground abyss.

Next, we assigned half of the meta-soil to damp the $X$-direction ground motion. The mass density matrix then takes the form $\rho_Z(\omega) = \rho_X(\omega) = 0.5\rho_0(f(\omega)+1)$. The transmission found is 0.90, higher than the case with all the meta-soil to damp the $Z$-direction vibrations. This is because the $Z$-component of the Rayleigh waves is significantly larger than the $X$-component. Therefore, in the remaining part of the paper we will use meta-soil and generic TMD's which only damp the $Z$-direction vibrations.

**(B)    Compare with conventional means**

For comparison, we have calculated three cases where conventional structures are used to block the seismic waves, namely the case with the empty wells (1$^{st}$ case), the case of wells filled with solid steel (2$^{nd}$ case), and the case of a continuous solid steel wall of 5 meters in thickness across the wave front (3$^{rd}$ case). The results are shown in the top, the middle, and the lower graph in Fig. 2(d). The empty-well case has the highest transmission and little absorption. The transmission of both the steel-well case and the continuous steel wall case have comparable apparent absorption of $A = 0.09$, which is due to the waves scattered down to the abyss rather than really being absorbed, as there is no dissipation mechanism in the structures. The transmission for case-1 is 0.99, i. e., the empty wells have almost no attenuation to the Rayleigh waves. The transmission for the steel wells is 0.95, and that for the continuous steel wall is 0.91. Both are less than the wells with meta-soil, showing clearly their advantage over conventional structures.

## (C)  Well length dependence of transmission

Figure 3(a) depicts the well length dependence of *t, r, A* for a single row of wells. All the scattering coefficients start to show sign of saturation for well length beyond 100 meters, indicating that the length of the wells should be comparable to the wavelength of the seismic waves. On the other hand, it is clear that a single row of wells, regardless of its length, can only attenuate the seismic wave amplitude by less than 20 %, which is clearly inadequate for earthquake protection.

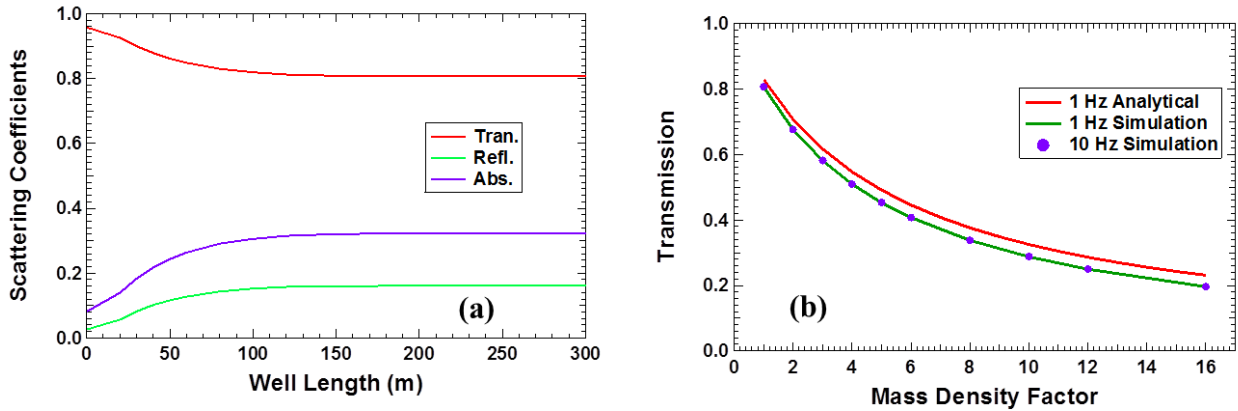

Figure 3  (a) Well length dependence of scattering coefficients at 1 Hz for wells filled with standard meta-soil mass density. (b) The mass density factor dependence of the 1 Hz transmission coefficient of wells 100 meter in length (solid curves). The green curve is the simulation results for wells filled with standard meta-soil with resonant frequency at 1 Hz, while the red one is from Eq. (6). The purple disks are the simulation results at 10 Hz with the meta-soil with 10 Hz resonant frequency and $0.1\rho_0$ in mass density.

## (D)  Mass density dependence of transmission

Figure 3(b) depicts the transmission of a single row of 300-meter length wells with increased density. A mass density factor of 5 in the figure means $\rho_Z(\omega) = 5\rho_0 f(\omega)$ in the dynamic mass density of the meta-soil, and so on, while $f(\omega)$ remains the same as in Eq. (1) with the resonant frequency of 1 Hz and $Q = 10$. The green curve is the transmission of 1 Hz seismic waves through the wells obtained by simulations. The red curve is obtained from Eq. 3(a), which agrees with the simulation results quite well. An important prediction of Eq. 3(a) is the scaling law that the transmission remains the same if the product of the mass density times the frequency remains constant. Indeed, when the frequency of the seismic waves and the resonant frequency of the meta-soil are both increased to 10 Hz, while reducing the mass density from $\rho_0$ to $0.1\rho_0$, the transmission curve as a function of mass density factor obtained from simulations (the purple points) agree perfectly with the 1 Hz one. The scaling law obtained from the simple analytic derivations is therefore verified by numerical simulations, and could be conveniently applied to estimate the attenuation of seismic waves at other frequencies. In particular, it brings out an important prediction that the attenuation of higher frequencies seismic waves requires lower mass density, or less number of dampers. The scaling law is the generalization of acoustics mass density law [20] in the seismic wave domain. Our local resonators for seismic wave attenuation are essentially a dynamic mass effect, and their resonances lead to mass magnification and maximum energy dissipation.

**(E)     Multiple rows of wells**

We now investigate the seismic wave transmission attenuation by multiple rows of wells. The interval between adjacent rows is fixed at 10 meters from center to center. Figure 4(a) depicts the transmission vs the number of rows of wells at different well length. The transmission attenuation increases with the increase of the row numbers without signs of saturation. For shallow wells with length ranging from 20 to 100 meters, the attenuation increases with well length, while beyond 100 meters in well length the increase in attenuation slows down considerably. The difference in attenuation between the 150-meter wells and the 200-meter wells is rather insignificant. This is consistent with the results shown in Fig. 3(a). In terms of cost effectiveness, the 100-meter wells are probably the most economical for the 8[th] magnitude earthquakes in the present case. The dashed curve in Fig. 4(a) is the simulation results of a single row of 100-meter wells with the equivalent areal mass density as the corresponding number of rows of 100-meter wells, i. e., a mass density of 10 $\rho_0$ corresponds to 10 wells with the standard mass density $\rho_0$. The attenuation by a single row of 10 $\rho_0$ of 100-meter wells is significantly inferior to that of the 10 rows of 100-meter wells with mass density $\rho_0$, indicating that deploying the dampers in multiple rows of wells can provide higher attenuation than packing all the dampers closely in a single row.

Another interesting result indicated in Fig. 4(a) is that the attenuation of a single row of wells 200 meters in length (~ 3 dB) is in fact lower than that of 10 rows of 20 meters in length (~ 6 dB). This is expected, as the amplitude of the Rayleigh waves is the largest near the ground surface. For the same number of wells, if more dampers are placed near the surface instead of deeper underground, more wave energy could be dissipated. Besides lower transmission coefficient, the reflection coefficient of the 10 rows of the 20-meter wells is also very low ($r$ = 0.02) as compared to $r$ = 0.16 for the single row of 200-meter wells. The wave intensity distributions of the two cases are shown in Fig. 4(b). Significant energy flow (on the right side of the arrow) into the abyss can be seen in the latter case. Similar phenomenon was predicted in Ref. 17.

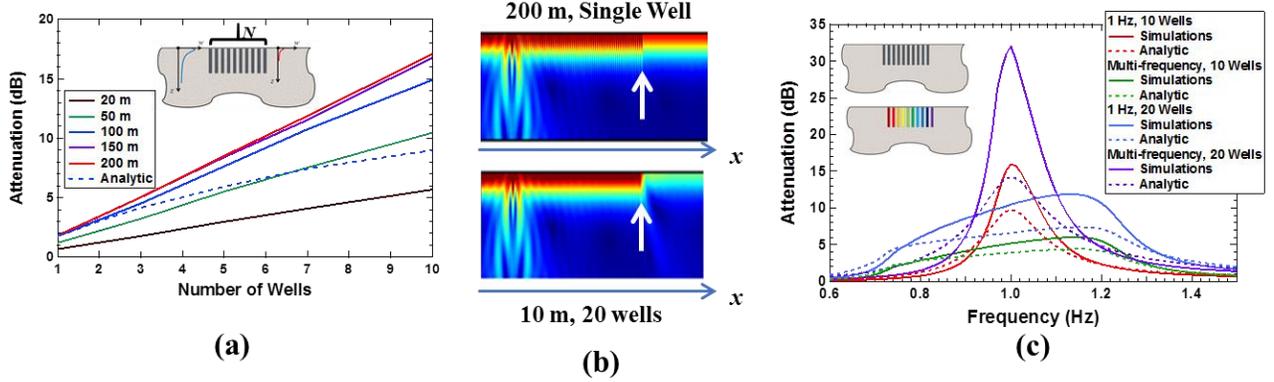

**(a)** **(b)** **(c)**

Figure 4  (a) The row number dependence of the transmission attenuation at 1 Hz with various well length; (b) The color coded wave intensity for a row of wells with 200 meter length (top) and 10 wells with 20 meter length (bottom); (c) The transmission attenuation spectra for several cases with meta-soli with different resonant frequencies and well arrangements.

The transmission attenuation spectra of several cases of multiple-row of wells are depicted in Fig. 4(c) as the solid curves, along with the analytic predictions using Eq. (3) in dashed curves of the same color. All the wells are 100 meters in length. The red solid curve is the attenuation of 10 rows of wells with the working frequency at 1 Hz. The width at half height is approximately 0.1 Hz, which is expected for $Q = 10$. The maximum attenuation of 16 dB occurs at 1 Hz. The red dashed curve is the analytic prediction assuming all the damper masses are projected onto the interface at $x = 0$. As is in the case in Fig. 4(a), projecting all the mass density onto a single row of wells would result in inferior transmission attenuation than spreading the it in multiple rows of wells. The green solid curve is the attenuation by 10 rows of wells with meta-soil of different working frequency for different rows. The working frequency $\omega_0$ in the first row is 0.75 Hz, and that of the last row is 1.20 Hz. The ones in between are in ascent order at 0.05 Hz increment. For the corresponding 20-row case each unit of two rows are of the same working frequency, and the working frequency of each unit follows the same ascent order as the 10-row case. The dashed curves with the same color as the solid curves are the corresponding analytic predictions obtained in the same way as for the first case. The 20 rows of wells with uniform working frequency of 1 Hz can provide more than 20 dB of transmission attenuation over a relatively narrow range, while the ones with distributed working frequency can provide broad band attenuation but the maximum attenuation is less than 20 dB. Therefore, dampers with multiple working frequencies are much desired to provide at least 20 dB of attenuation over a wide enough frequency range, and the DMR-like dampers with multiple mass blocks are good candidates. It is also seen that the analytical predictions under estimate the attenuation by a large margin for multiple rows of wells, so its effectiveness is limited to single row cases only.

**Perfect absorbers**

The above cases demonstrate good seismic wave barriers, but reflect considerable amount of wave back. Taking reference of the low reflection by the shallow wells, we designed a tiered array of wells in a wedge formation as shown in the insert of Fig. 5(a) to achieve total absorption. The first 9 rows of wells are, in consecutive order, 10, 20,,, 90 meters in length, while the last 5 rows are 100 meters in length. All the wells are filled with meta-soil with the resonant frequency $\omega_0 = 1$ Hz. The reflection, transmission, and absorption coefficients are depicted in Fig. 5(a) as the red, the green, and the blue curves. The reflection is below 0.04 throughout the frequency range of interest. The maximum transmission attenuation is 17.7 dB. The transmission attenuation would be over 20 dB if two more rows of 100-meter wells are added at the back of the wedge. The absorption peak value is 0.982. The reflection, transmission, and absorption coefficients for the incident waves onto the opposite side (the back side) of the wedge are shown in Fig. 5(a) as the disks with the corresponding colors. The front and the back side transmission spectra are the same within computational errors. This is expected according to wave propagation reciprocity. The front and the back side reflection spectra are about the same except for the resonant frequency region from 0.88 Hz to 1.28 Hz. The maximum difference, or asymmetry in reflection, is 0.18. The wedge is therefore a good candidate for large Willis coupling solids [27 − 29] at very low frequency.

To further validate the excellent seismic wave absorption of the wedge structure, we use discrete fictive TMD's to replace the meta-soil in the wells, as shown in the middle insert in Fig. 5(b). Limited by the available computation resources, we replace many DMR-like dampers, which have been homogenized as meta-soil, within a well segment 2 meters in height and 5 meter in diameter by a single generic TMD filling up the space. The block oscillator mass $M_B$ is equal to the meta-soil mass in the segment with $M_B = \pi r^2 h \cdot \rho_0 = 3.9 \times 10^4$ kg. The equivalent stiffness of the spring is taken as $k_s = M_B \omega_0^2$ and the equivalent viscosity of the dashpot is $\gamma_s = \dfrac{M_B \omega_0}{Q}$, so that the TMD has the same mass and response function as the meta-soil segment it represents. The anchor plate of the TMD is fixed on the wall of the well, and its mass is assumed to be negligible.

One issue to confirm is whether at resonance the local resonators are vibrating out of phase relative to the ground motion, which serves as the drive to the local resonators. This is verified by examining a fictive TMD placed on the soil surface. The results show clearly that the motion of the mass block is 90 degrees out of phase to the motion of the anchor, which corresponds to maximum dissipation of the soil vibration energy by the local resonator.

The transmission, reflection, and absorption spectra for the front side incidence are shown in Fig. 5(b) as the red, the green, and the purple curves. The reflection spectrum so obtained is much the same as the one obtained from the meta-soil case. Both show that the reflection is near zero around 1.12 Hz. The maximum transmission attenuation is 23.8 dB. The absorption maximum is 0.994.

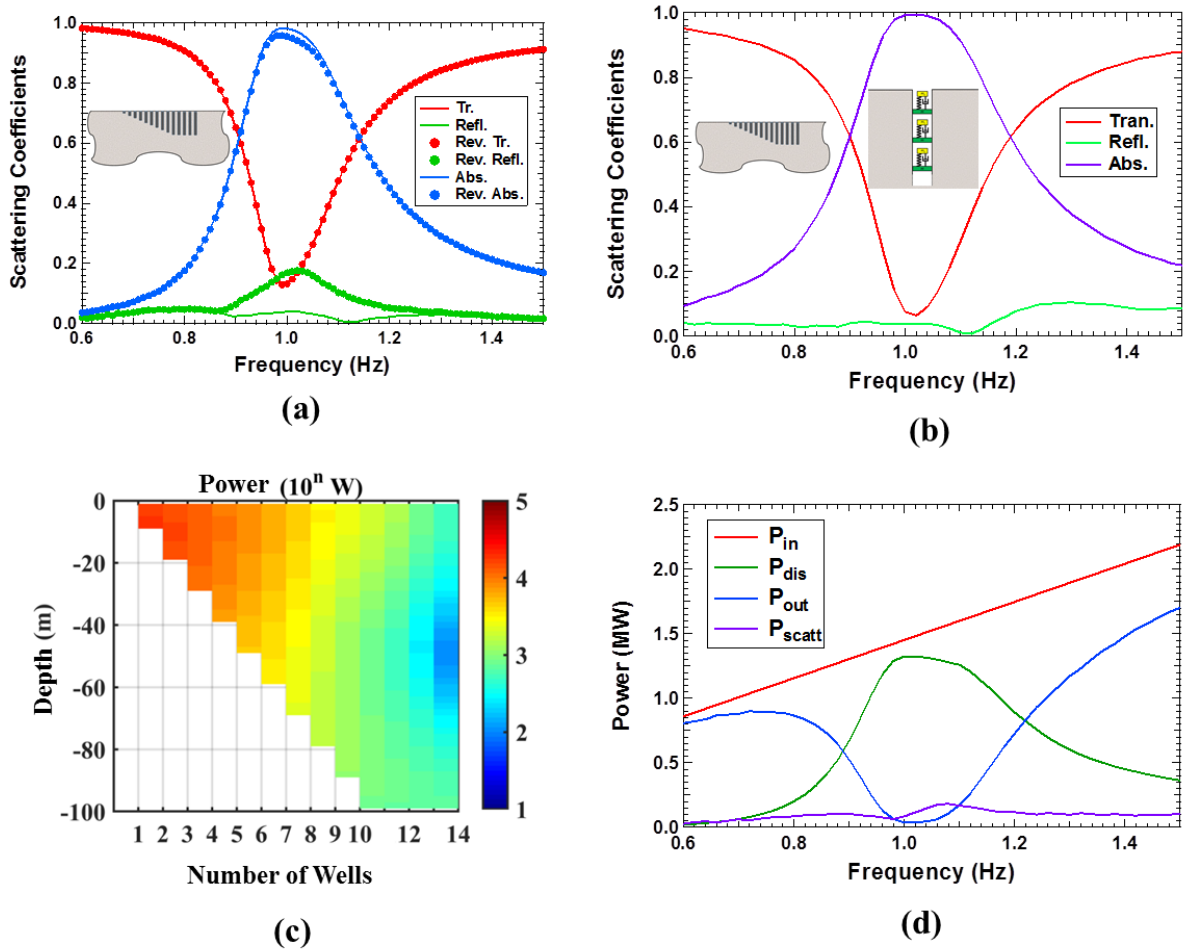

**(a)**

**(b)**

**(c)**

**(d)**

Figure 5  (a) The scattering coefficient spectra of the tiered well configuration. The solid curves are for the scattering coefficients when the waves are incident onto the short well side (front side), while the disks are the corresponding ones when the waves are incident from the back side. (b) The same coefficients spectra as in (a) but with the meta-soil replaced by generic TMD's as schematically shown in the middle insert. (c) The color coded dissipation power of each TMD; (d) The input power (red curve), the total dissipation power (green curve), the transmitted power (blue curve), and the reflected power (purple) spectra for the same case as in (c).

The discrete TMD's makes it possible to examine the power dissipation by each TMD. The color code dissipation map is shown in Fig. 5(c). The TMD's in the first row wells are dissipating power of the order of 100 KW. The seismic waves continues to penetrate and move around the bottom of the wedge, such that even the TMD's near the soil surface and the wedge bottom in the last row are still dissipating KW of power. Only the TMD's in the middle part of the last row remain almost dormant. As the total power dissipated by all the TMD's can now be counted, we can determine the wave energy scattered into the underground abyss. Figure 5(d) depicts the ratio of power of each process. At 1 Hz, over 91 % of the incident wave energy is dissipated by the dampers, 2.7 % is radiated as transmission and reflection waves, and 6 % is scattered to abyss.

The discrete TMD's used in this case are probably still too massive to satisfy criterio-4. Their relatively large size increases the local mass density mismatch that leads to a slight over estimation of the transmission attenuation than the meta-soil. For more realistic investigations of deep subwavelength dampers the meta-soil approach is preferred.

**Summary and Discussions**

We have taken a major step beyond proof of concept in seismic wave attenuation by metamaterials. First, we have proposed four major criteria for seismic barriers to attenuate the highest (8[th]) magnitude earthquakes recorded in human history. Second, we have shown that all four criteria can be met by using the compact DMR-like local resonators using folded steel sheets to replace the membrane in a conventional DMR, which are homogenized into meta-soil because of their deep subwavelength dimension scale. Using wells dug into the ground filled with the meta-soil, we have demonstrated numerically a number of cases where sufficient transmission attenuation (20 dB or above) could be realized. We have also shown an array of the wells in wedge form that could absorb over 98% of the incident Rayleigh waves, while providing 17.7 dB of transmission attenuation. With multiple working frequency dampers, a seismic absorption barrier that satisfies both criserio-1 (broadband) and criterio-4 (20 dB attenuation) are within reach.

The main mechanism of the seismic wave attenuation by the local resonators is the scattering by the resonant dynamic mass amplification and absorption by the resonant dissipation. The deploy­ment of the local resonators could be discontinuous and irregular at locations where large bedrock would provide natural obstacles. The combined effects of local resonators are mostly cumulative. This is expected as each local resonator absorbs and scatters seismic waves when excited, independent of the presence of the others. For protection of individual buildings, one can design specific frequency of maximum attenuation that matches the first resonant frequency of the building that block not only the Rayleigh waves, but also the pressure and the shear waves when the local resonators are deployed around the bases of buildings. The scaling law will be particularly handy for designs of local resonators against higher frequency ground motions excited by train or heavy vehicle movements. The results reported here can serve as a reference for other deep subwavelength locally resonant vibration damping devices placed underground or on the ground surface as long as they can be represented by generic compact TMD's and homogenized into meta-soil.

**Supplementary Information**

A) Generic tuned mass dampers

Consider a generic tuned mass damper (TMD) consisting of single oscillator with mass $m$ attached to a spring and a dashpot in parallel, as shown in Fig. 1(a) in the main text. The other ends of the spring and the dashpot are attached to the anchor which is oscillating with amplitude $A$ and at frequency $\omega$. The force exerted on the anchor by the spring and the dashpot is $F = -kx - i\gamma\omega x$, where $x$ is the displacement of the oscillator relative to the anchor, $k$ is the force constant of the spring, and $\gamma$ is the viscous coefficient. The effective mass, defined as $\tilde{m} \equiv -\dfrac{F_{Ext}}{A\omega^2} = m_A - \dfrac{k + i\gamma\omega}{\omega^2}\dfrac{x}{A}$, where $m_A$ is the mass of the anchor and $x$ is the relative distance between the anchor and the oscillator, can be readily obtained by solving the dynamic equation of the oscillator. The equation of motion of the oscillator, taking the anchor as the reference, is $\omega^2 Am - kx - i\gamma\omega x = -m\omega^2 x$, where the term $\omega^2 Am$ is the inertia force due to the fact that the anchor is oscillating. Then, using the conventional definition $\dfrac{\gamma}{m} \equiv \dfrac{\omega_0}{Q}$ and $k \equiv m\omega_0^2$, we have

$$\tilde{m} = m_A + \frac{\omega_0^2 + i\omega\omega_0/Q}{\omega_0^2 - \omega^2 + i\omega\omega_0/Q}m \qquad \text{(S-1)}$$

## B) The DMR-like TMD's

The original rubber membrane in a DMR [23] is replaced by a stainless steel sheet about 0.1 mm in thickness alternating folded into a zigzag pattern. The width of each zig and zag is about 10 mm, and the length of the folded sheet in loose state is about half of its original length of the flat sheet. The folded form of the sheet greatly softens the restoring forces of the flat sheet, such that a block mass about 2 g can already lead to a resonant frequency of 10 Hz. The folded corners are the regions of high elastic energy concentration because of the large curvature that provide much stronger energy dissipation than the flat sheets. That is the reason why the $Q$-factor of the DMR-like damper in Fig. 1(c) at 10 Hz is only 30, as compared to over 1000 if flat sheets are used.

The free vibration of the damper was measured by mounting it, via its anchor, on a force sensor. After an initial gentle knock on the block mass (the nut), the free vibration data were then measured by the force sensor underneath. The time domain data were then Fourier transformed into the vibration spectrum.

## References

[1]    Roberto, V. *Fundamental Concepts of Earthquake Engineering* (CRC Press, 2009).

[2]    Martinez-Sala, R., Sancho, J., Sanchez, J. V., Gomez, V., Llinares, J. & Meseguer, F. Sound attenuation by sculpture. *Nature* **378**, 241 (1995).

[3]     Meseguer, F., Holgado, M., Caballero, D., Benaches, N., Sánchez-Dehesa, J., López, C., & Llinares, J. Rayleigh-wave attenuation by a semi-infinite two-dimensional elastic-band-gap crystal. *Phys. Rev. B* **59** (19): 12169 (1999).

[4]     Brûlé, S., Javelaud, E. H., Enoch, S. & Guenneau, S. Experiments on Seismic Metamaterials: Molding Surface Waves. *Phys. Rev. Lett.* **112**, 133901 (2014).

[5]     Du, Q., Zeng, Y., Huang, G. & Yang, H. Elastic metamaterial-based seismic shield for both Lamb and surface waves. *AIP Advances* **7**, 075015 (2017).

[6]     Liu, Z., Zhang, X., Mao, Y., Zhu, Y., Yang, Z., Chan, C. & Sheng, P. Locally resonant sonic materials. *Science* **289**, 1734 (2000).

[7]     Finocchio, G., Casablanca, O., Ricciardi, G., Alibrandi, U., Garescì, F., Chiappini, M. & Azzerboni, B. Seismic metamaterials based on isochronous mechanical oscillators. *Appl. Phys. Lett.* **104**, 191903 (2014).

[8]     Krödel, S., Thoméa, N. & Daraio, C. Wide band-gap seismic metastructures. *Extreme Mechanics Letters* **4**, 111–117 (2015).

[9]     Achaoui, Y., Ungureanu, B., Enoch, S., Brûlé, S. & Guenneau, S. Seismic waves damping with arrays of inertial resonators. *Extreme Mechanics Letters* **8**: 30–37 (2016).

[10]    Miniaci, M., Krushynska, A., Bosia, F. and Pugno, N. M. Large scale mechanical metamaterials as seismic shields. *New J. Phys.* **18**, 083041 (2016).

[11]    Palermo, A., Krödel, S., Marzani, A. & Daraio, C. Engineered metabarrier as shield from seismic surface waves. *Sci. Rep.* **6** (19238):1–7 (2016).

[12]    Colombi, A., Roux, P., Guenneau, S., Gueguen, P. & Craster, R. V. Forests as a natural seismic metamaterial: Rayleigh wave bandgaps induced by local resonances. *Sci. Rep.* **6** (19238):1–7 (2016).

[13]    Colquitt, D. J., Colombi, A., Craster, R. V., Roux, P. & Guenneau, S. R. L. Seismic metasurfaces: Sub-wavelength resonators and Rayleigh wave interaction. *J. Mech. Phys. Solids* **99**, 379–393 (2017).

[14]    Palermo, A., Vitali, M. & Marzani, A. Metabarriers with multi-mass locally resonating units for broad band Rayleigh waves attenuation. *Soil Dynamics and Earthquake Engineering* **113**, 265 – 277 (2018).

[15]    Zeng, Y. Xu, Y., Deng, K., Zeng, Z., Yang, H., Muzamil, M. & Du, Q. Low-frequency broadband seismic metamaterial using I-shaped pillars in a half-space. *Journal of Applied Physics* **123**, 214901 (2018).

[16]    Lim, C. W. & Reddy, J. N. Built-up structural steel sections as seismic metamaterials for surface wave attenuation with low frequency wide bandgap in layered soil medium. *Engineering Structures* **188**, 440–451 (2019).

[17]    Colombi, A., Colquitt, D., Roux, P., Guenneau, S. & Craster, R. V. A seismic metamaterial: the resonant metawedge. *Sci. Rep.* **6** (27717):1–6. (2016).

[18]    Woods, R. D. Screening of Surface Waves in Soils, Technical Report No. IP-804, University of Michigan (1968).

[19]  Xu, Y., Yang, Z. & Cao, L. Deflecting Rayleigh surface acoustic waves by a meta-ridge with a gradient phase shift. *J. Phys. D: Appl. Phys*. **51**: 175106 (2018).

[20]  Brekhovskikh, L. M. *Waves in Layered Media* (Academic Press, New York, 1980) 2[nd] edition.

[21]  Ormondroyd, J. & Den Hartog, J. P. The theory of dynamic vibration absorber. *Trans. Am. Soc. Mech. Eng*. **50**: 9–22 (1928).

[22]  Den Hartog, J. P. *Mechanical vibrations* (Courier Corporation, 1985).

[23]  Yang, Z., Yang, M., Chan, N. & Sheng, P. Membrane-type acoustic metamaterial with negative dynamic mass. *Phys. Rev. Lett.* **101**, 204301 (2008).

[24]  Sun, L., Au-Yeung, K., Yang, M., Tang, S., Yang, Z. & Sheng, P. Membrane-type resonator as an effective miniaturized tuned vibration mass damper. *AIP Advances* **6**(8), 085212 (2016).

[25]  Au Yeung, K. Y. Yang, B. Sun, S. Bai, K. Yang, Z. Super Damping of Mechanical Vibrations, arXiv:1811.08621

[26]  Sun, L. & Yang, Z. Experimental investigation of vibration damper composed of acoustic metamaterials, *Applied Acoustics* **119**, 101–107 (2017).

[27]  Willis, J. R. Variational principles for dynamic problems for inhomogeneous elastic media, *Wave Motion* **3**, 1 (1981).

[28]  Milton, G. W. New metamaterials with macroscopic behavior outside that of continuum elastodynamics, *New Journal of Physics* **9**, 359 (2007).

[29]  Yongquan Liu, Zixian Liang, Jian Zhu, Lingbo Xia, Olivier Mondain-Monval, Thomas Brunet, Andrea Alù, and Jensen Li, Willis Metamaterial on a Structured Beam, *Phys. Rev. X* **9**, 011040 (2019)